\documentclass[paper,12pt]{article}
\pdfoutput=1

\usepackage{jcappub}
\usepackage{ulem}

\definecolor{agreen}{rgb}{0.1000,0.4000,0.1000}
\definecolor{ared}{rgb}{0.7000,0.1000,0.1000}

\title{A FLUKA Study of $\beta$-delayed Neutron Emission for the Ton-sized 
                      DarkSide Dark Matter Detector}

\author[1]{A. Empl\note{Corresponding author.}}
\author[]{and E. V. Hungerford}

\affiliation[]{Department of Physics, University of Houston, Houston, TX 77204}
\emailAdd{aempl@central.uh.edu}

\abstract{
 In the published cosmogenic background study for a ton-sized DarkSide
 dark matter search, only prompt neutron backgrounds coincident with
 cosmogenic muons or muon induced showers were considered, although
 observation of the initiating particle(s) was not required. The present
 paper now reports an initial investigation of the magnitude of cosmogenic
 background from $\beta$-delayed neutron emission produced by cosmogenic
 activity in DarkSide.  The study finds a background rate for
 $\beta$-delayed neutrons in the fiducial volume of the detector on the
 order of $<$ 0.1 event/year.  However, detailed studies are required to
 obtain more precise estimates.  The result should be compared to a
 radiogenic background event rate from the PMTs inside the DarkSide
 liquid scintillator veto of 0.2 events/year.
}


\begin{document}
\maketitle
\flushbottom

\section*{$\beta$-decay delayed neutrons}

 Most neutrons emitted by nuclear de-excitation after cosmogenic activation
 were included in our original investigation~\cite{empl}.  Such emission is
 governed by strong interactions and hence they occur in time coincidence
 with the initiating muon or muon induced shower, although the simulation did
 not required the initiating particle to be observed. This study reported the
 response of the DarkSide (DS) water (CTF) and scintillator (LSV) vetoes to
 the number of all such neutrons.

 \vskip 3mm \hskip -\parindent
 However it is also possible that radioactive nuclear isotopes produced by
 cosmogenic activation undergo $\beta$-decay to a daughter nucleus which
 itself may then promptly emit a neutron.
 Such decays are not in time coincidence with an incident muon or 
 muon induced shower and would not have been included in our original study. 
 The study did include neutrons produced in materials (including the cavern
 walls) with or without a charged particle incident in the vetoes.  However
 the number of these neutrons is severely attenuated by shielding before
 entering the active volume of the DS dark matter detector.  This is opposed
 to neutron emission from delayed $\beta$-activity produced in or near the
 active detector volume.

\begin{figure}[htb]
    \vskip -1mm
    \centering
    \includegraphics[width=.745\textwidth]{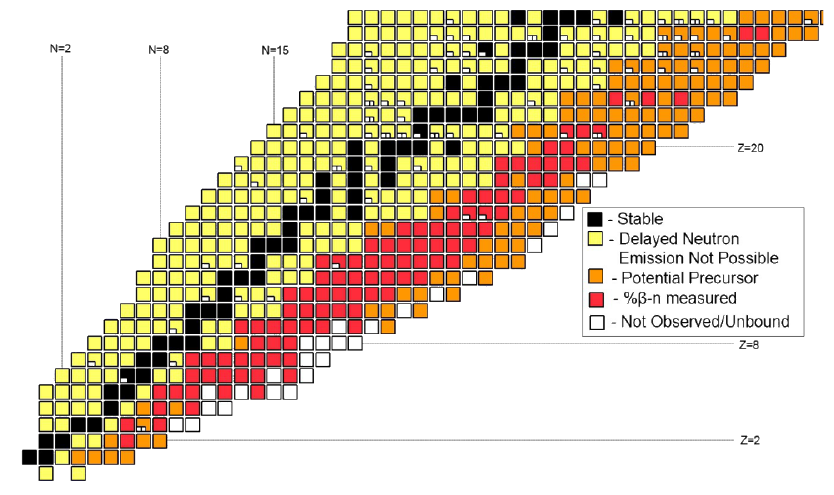}
    \caption{ Potential Precursors in the A $\le$ 72 Region~\cite{birch1}.}
   \label{NvsP}
   \end{figure}

 \vskip 3mm \hskip -\parindent
 The topic of delayed neutrons has been studied extensively, for example in
 the context of nuclear reactors and the R-process in astronomy.  The graph
 shown in Figure~\ref{NvsP} was taken from a report on an ongoing effort to
 prepare an evaluated database for delayed neutron emission~\cite{birch1}.  
 The possible precursor nuclei which can result in delayed neutron emission
 are indicated by the red (measured) and orange (model predicted) boxes in
 the figure as a function of proton and neutron numbers.\\

\begin{figure}[htb]
    \centering
    \includegraphics[width=.425\textwidth]{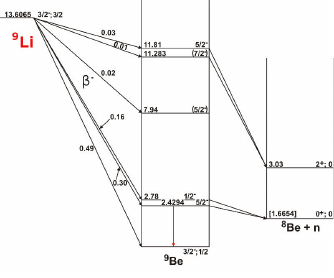} \hfill
    \includegraphics[width=.565\textwidth]{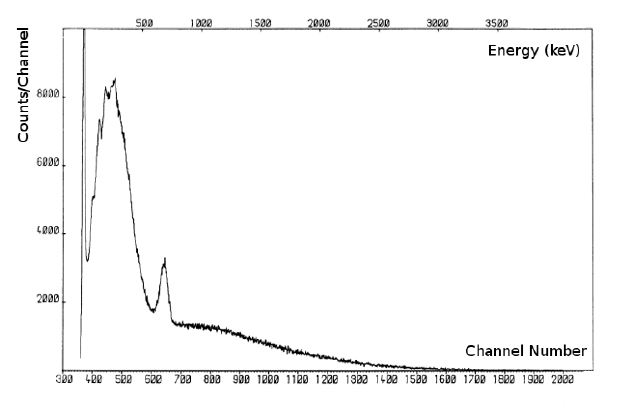}
    \caption{ left) Level scheme for $^{9}$Li precursor isotope leading sometimes
               to delayed neutron emission by the creation of $^{8}$Be.
                \,\, right) Measured neutron kinetic energy spectrum for delayed
                 neutrons from the $^{9}$Li precursor. The visible small peak
                  corresponds to the most likely transition at dE= 0.682 MeV
                   (= 0.764 MeV in Table~\ref{fract}). }
   \label{9Li}
   \end{figure}

 Further, one can find experimental information on nuclear levels for some
 precursor nuclei, their excited daughter isotopes and the resulting isotope
 after neutron emission~\cite{tunl}.  As an example, the level diagram for
 the $^{9}$Li precursor which can lead to $^{8}$Be + n is shown on the left
 in Figure~\ref{9Li}.

 \mbox{} \vskip -8mm \hskip -\parindent
 Table~\ref{fract} gives the decays and probabilities for delayed neutron
 emission from different levels in the $^{8}$He, $^{9}$Li, $^{13}$B and $^{17}$N
 precursor nuclei.  Only transitions with at least 0.1\% probability and no
 additional emitted charged particles are included.  These four isotopes
 comprise 95\% of the produced precursor nuclei obtained from a FLUKA
 simulation for the DS detectors.

 \begin{table}[htb]
   \vskip 4mm
   \small
   \centering
   \begin{tabular}{l|rrc}
 $^{8}$He $\rightarrow$ $^{7}$Li+n & levels  (MeV) &    dE$_{n}$ (MeV)   &  fraction (\%)  \\
     \hline &&&\\[-10pt]
      &    5.40   - 0.4776    &    4.82             &  $\sim$ 8  \\
      &    3.21   - 0.4776    &    7.32             &  $\sim$ 8  \\
      &&&\\[-8PT]
 $^{9}$Li $\rightarrow$ $^{8}$Be+n & levels  (MeV) &    dE$_{n}$ (MeV)   &  fraction (\%)  \\
     \hline &&&\\[-10pt]
      &    11.81   - 3.03     &    8.78             &   3  \\
      &    11.283  - 3.03     &    8.253            &   1  \\
      &     2.78   - 1.6654   &    1.114            &  16  \\
      &     2.4294 - 1.6654   &    0.764            &  30  \\
      &&&\\[-8PT]
 $^{13}$B $\rightarrow$ $^{12}$C+n & levels  (MeV) &    dE$_{n}$ (MeV)   &  fraction (\%)  \\
     \hline &&&\\[-10pt]
      &    8.860   - 4.9463   &    3.914            &   0.2  \\
      &&&\\[-8PT]
 $^{17}$N $\rightarrow$ $^{16}$O+n & levels  (MeV) &    dE$_{n}$ (MeV)   &  fraction (\%)  \\
     \hline &&&\\[-10pt]
      &    5.939   - 4.143    &    1.796            &   7.4   \\
      &    5.379   - 4.143    &    1.236            &   50    \\
      &    5.085   - 4.143    &    0.942            &   0.6   \\
      &    4.554   - 4.143    &    0.441            &   38    \\
    \end{tabular}
   \caption{Level schemes for delayed neutron emission of the most copiously
              produced precursor nuclei.  Only the sum fraction of 16\% is
                  available for $^{8}$He. }
    \label{fract}
    \end{table}

 In the case of delayed neutron production from the $^{9}$Li precursor, an
 experimental measurement of the emitted neutron kinetic energy spectrum is also
 available~\cite{numan}.  This is shown on the right in Figure~\ref{9Li}.  
 The peak near
 680 keV corresponds to the most frequent neutron energy from the decay of the 
 lowest excited states of $^{9}$Be to $^{8}$Be + n, but the experimental value 
 does not quite match the expected dE= 764 keV.  This emission is expected in
 30\% of the $^{9}$Li decays.

\section*{Precursor production in FLUKA}

 Good agreement is obtained between the experimental results of
 Borexino~\cite{bxcos} and a FLUKA simulation of cosmogenic
 isotope production in light target materials, ie liquid scintillator.
 For example, the predicted and measured yield for the $^{9}$Li precursor 
 are  3.1$\pm$0.4  and  2.9$\pm$0.3
 \, $[\times 10^{-7}\,(\mu\,\textrm{g}/\textrm{cm}^2)^{-1}]$ respectively.
 Even though FLUKA~\cite{fluka1,fluka2} does include the treatment of the
 decays of radioactive nuclei, it does not correctly treat the subsequent
 delayed emission of neutrons~\cite{fluka3}.  Thus a list of precursor
 nuclei and their production properties, location, energy, and time, were
 recorded in a FLUKA simulation for the DS ton-sized experiment.
 The number of simulated events corresponds to a lifetime of
 approximately 34 years.  Statistical uncertainties of the simulated
 results are small ($<$ 10\%) when compared to systematics which are not
 reported.

\begin{figure}[bht]
    \centering
    \includegraphics[width=.320\textwidth]{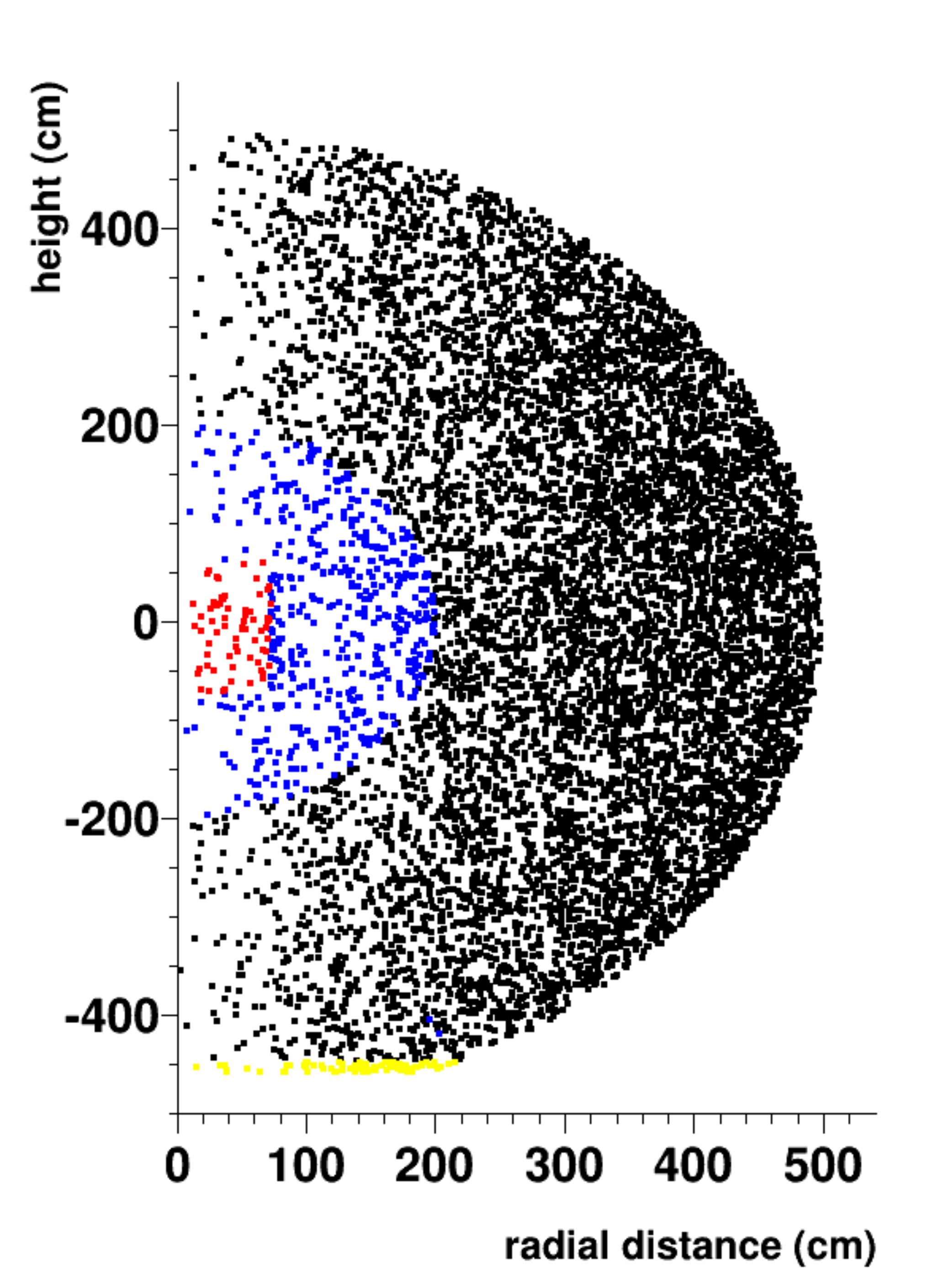}
    \includegraphics[width=.670\textwidth]{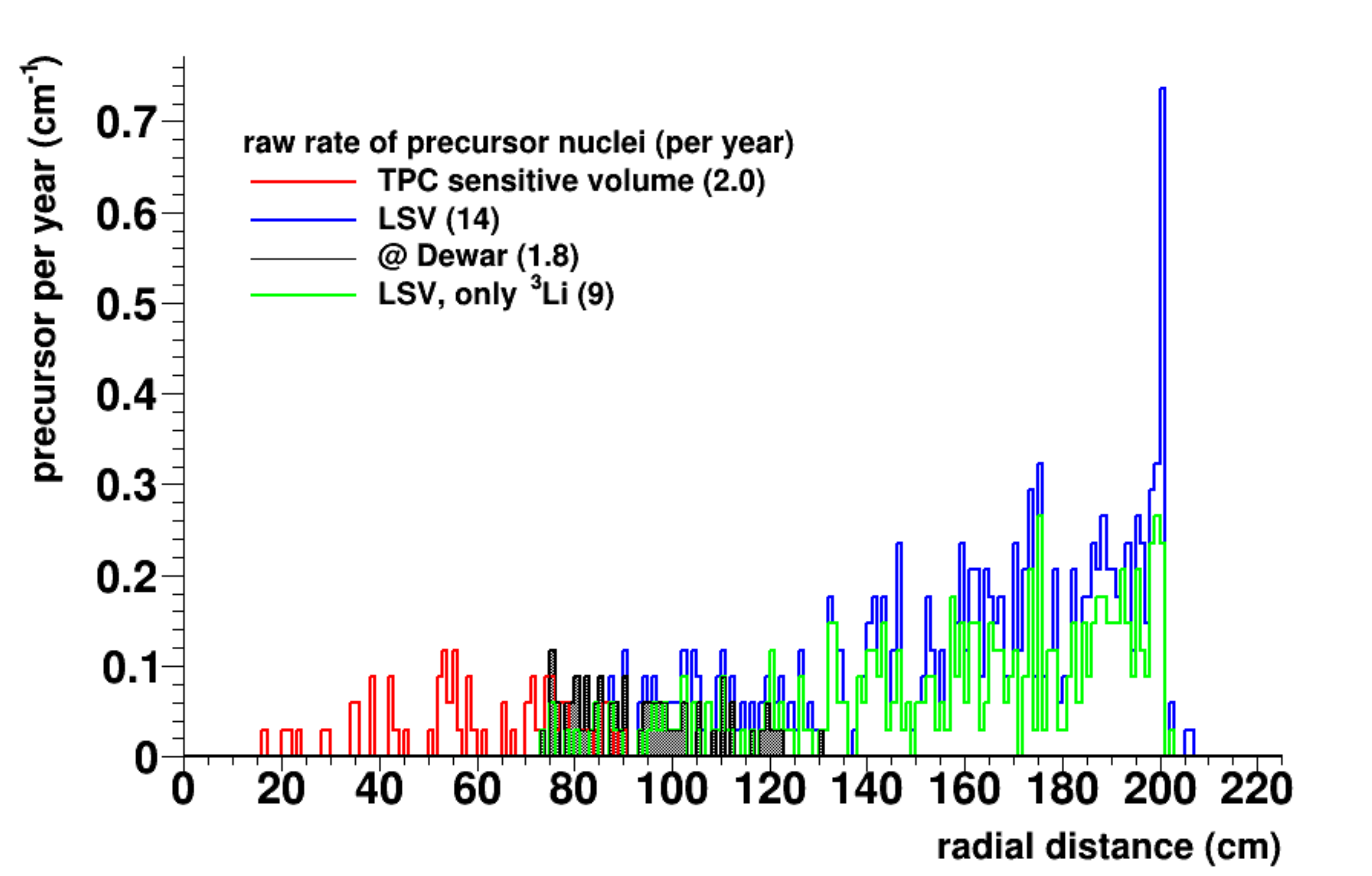}
    \caption{ left) Location of precursor production in the DarkSide setup
               given as a function of radial distance versus height; red -
                sensitive volume, blue - LSV and black - CTF (R$<$500~cm,
                 yellow symbols correspond to the CTF steel floor support). \,\,
             right) Precursor production as function of radial distance from
              center of sensitive volume; red - sensitive volume, blue - LSV and
                green - $^{9}$Li precursors inside LSV.  Precursor nuclei
                 created at and inside the Dewar are shown by the black hatched
                  region and the spike at 200 cm indicates production in the LSV
                   vessel material. }
   \label{ds}
   \end{figure}

 \vskip 3mm \hskip -\parindent
 On the left in Figure~\ref{ds} the positions of all precursor nuclei are
 shown for distances $\le$~500~cm from the center of the sensitive volume.
 The blue and red symbols indicate production within the LSV and
 inner sensitive volume respectively, while the black symbols
 correspond to locations fully contained inside the CTF.  It is assumed
 that delayed neutrons produced in the CTF do not result
 in background due to their distance from the sensitive volume and moderation
 in the water and scintillator vetoes.  Delayed neutrons created inside
 the liquid argon sensitive volume are also rejected since the coincident
 $\beta$-decay electron and possible charged particle emissions would be
 detected with very high probability.

 \vskip 3mm \hskip -\parindent
 The raw rate of precursor nuclei production as a function of distance from
 the center of the sensitive volume is given on the right in Figure~\ref{ds}.
 The blue solid histogram corresponds to the per year production rate in
 the LSV while the green solid histogram shows the production fraction from
 the $^{9}$Li precursor.  More critical precursor nuclei created in the
 passive material inside the Dewar close to the sensitive volume are indicated by
 the hatched black region.  Their total predicted rate of 1.8 events per year
 can be compared to the expected rate of radiogenic neutrons with comparable
 energies produced by the DS photomultipliers on the inner detector which
 is 56 (=\,556\,$\times$\,0.1) neutrons per year~\cite{dsG2}.

 \vskip 3mm \hskip -\parindent
 The total rate of precursor production inside the full LSV is approximately
 14.3 per year. Table~\ref{fract} lists the yearly production rates for the
 four most frequent precursor nuclei.  The expected rate of delayed neutron
 production from these precursor nuclei is estimated by applying the
 transition probabilities as given by the level schemes.  $^{9}$Li is by
 far the most frequently produced precursor nucleus in the DS experiment.  In
 50\% of the $^{9}$Li decays a delayed neutron will be produced and of those
 60\% originate from the transition between the lowest excited level in
 $^{9}$Be to $^{8}$Be + n with a rather small energy difference of dE=764~keV.

 \begin{table}[htb]
   \vskip 2mm
   \small
   \centering
   \begin{tabular}{l|c|c|c|c|c}
    precursor nuclei  & $^{8}$He  &  $^{9}$Li  &  $^{13}$B   & $^{17}$N  &  of total  \\
    \hline &&&&&\\[-10pt]
     fraction (\%)    &      8     &    63        &    19    &  7   &  95/100     \\
     events (y$^{-1}$)    & \,\, 1.15 \,\, & \,\, 8.9 \,\,  & \,\, 2.6 \,\,  &  \,\, 1.0 \,\, & 13.6/14.3  \\
     delayed neutrons (y$^{-1}$)     &   0.18  &   4.5  &  0.054  & 0.95  &          \\
    \end{tabular}
   \caption{FLUKA prediction for the production of the four most frequently found
             precursor nuclei in the DS experiment. }
    \label{fract}
    \end{table}

 \section*{Backgrounds from $\beta$-delayed neutrons in DarkSide}

 To complete the simulation, neutrons found from the four most frequently
 produced precursor nuclei were then injected in the FLUKA simulation
 at their respective positions, energies, and rates as described above.
 FLUKA then provided an estimate of the number of these neutrons which
 reach the sensitive volume along with the energy deposited
 into both the sensitive region and the LSV.

 \vskip 3mm \hskip -\parindent
 Events with raw energy deposition between 10~keV\,$<$\,dE\,$<$\,1~MeV in
 the sensitive volume are conservatively considered potential background to
 DS~\cite{empl}.  The rate of background events from $\beta$-delayed
 neutrons is given in Table~\ref{res} for the four most copiously produced
 precursor nuclei.  Neutrons from other FLUKA predicted precursor nuclei
 would add no more than 10\% to this result.  A total of approximately 2.3
 raw background events per year are predicted.  Note that 97\% of these
 events originate from precursor nuclei created in the Dewar.  Only a
 fraction of these events should give rise to a WIMP like signature for
 the DS dual phase liquid argon time projection chamber.  Thus a study with
 detailed detector response is required to address a more realistic rate
 prediction.

 \begin{table}[htb]
   \vskip 2mm
   \small
   \centering
   \begin{tabular}{l|c|c|c|c}
    precursor nuclei  & $^{8}$He & $^{9}$Li & $^{13}$B  & $^{17}$N  \\
    \hline &&&&\\[-10pt]
     events (y$^{-1}$)  & \,\,  0.003 \,\, & \,\, 1.83 \,\, & \,\, 0.02 \,\, & \,\, 0.42 \,\, \\
     with no LSV signal (y$^{-1}$)  & 4$\times10^{-5}$ & 0.07 & 5$\times10^{-4}$  & 0.01 \\
    \end{tabular}
   \caption{FLUKA predicted rate of delayed cosmogenic background events for
             the DS experiment.  See text for the definition of signals
              for the inner sensitive detector volume and the LSV. }
    \label{res}
    \end{table}

 Most of these events will not be vetoed by the DS outer detectors since
 the $\beta$-decay delayed neutrons occur well after the originating
 cosmogenic muon or shower.  However decay signals in the LSV and CTF are
 recorded and are available for offline analysis where events can be removed
 using a conservative lower limit of dE$_{LSV}$\,$>$\,1~MeV for the raw
 energy deposited in the LSV.  This is shown in Table~\ref{res} where a
 total rate of $<$\,~0.1 background events per year to the DS experiment
 from $\beta$-delayed cosmogenic neutrons is found.

 \section*{Conclusion}

 The impact of cosmogenic $\beta$-decay delayed neutron background in the
 DarkSide ton-sized experiment was studied. Even with very conservative
 assumptions a rate of $<$~0.1 events per year is found.  Fluka predicts
 approximately 97\% of the delayed neutron background originates from
 precursor nuclei created in the DS Dewar.  According to the simulation,
 the production rate of these precursor nuclei inside the Dewar is 1.8
 per year.  This should be compared to expected 56 neutrons per
 year from DS inner detector PMTs.  Thus the background contribution from
 cosmogenic $\beta$-decay delayed neutrons is small, but to improve the
 prediction, a careful simulation of the  detailed detector response is
 required.

  \section*{Appendix}

 Level scheme for daughter nuclei after $\beta$-decay from three
 precursor, $^{8}$He, $^{13}$B and $^{17}$N, and resulting
 isotopes after neutron emission (from reference~\cite{tunl}).

   \begin{figure}[htb]
     \centering
     \includegraphics[width=.95\textwidth]{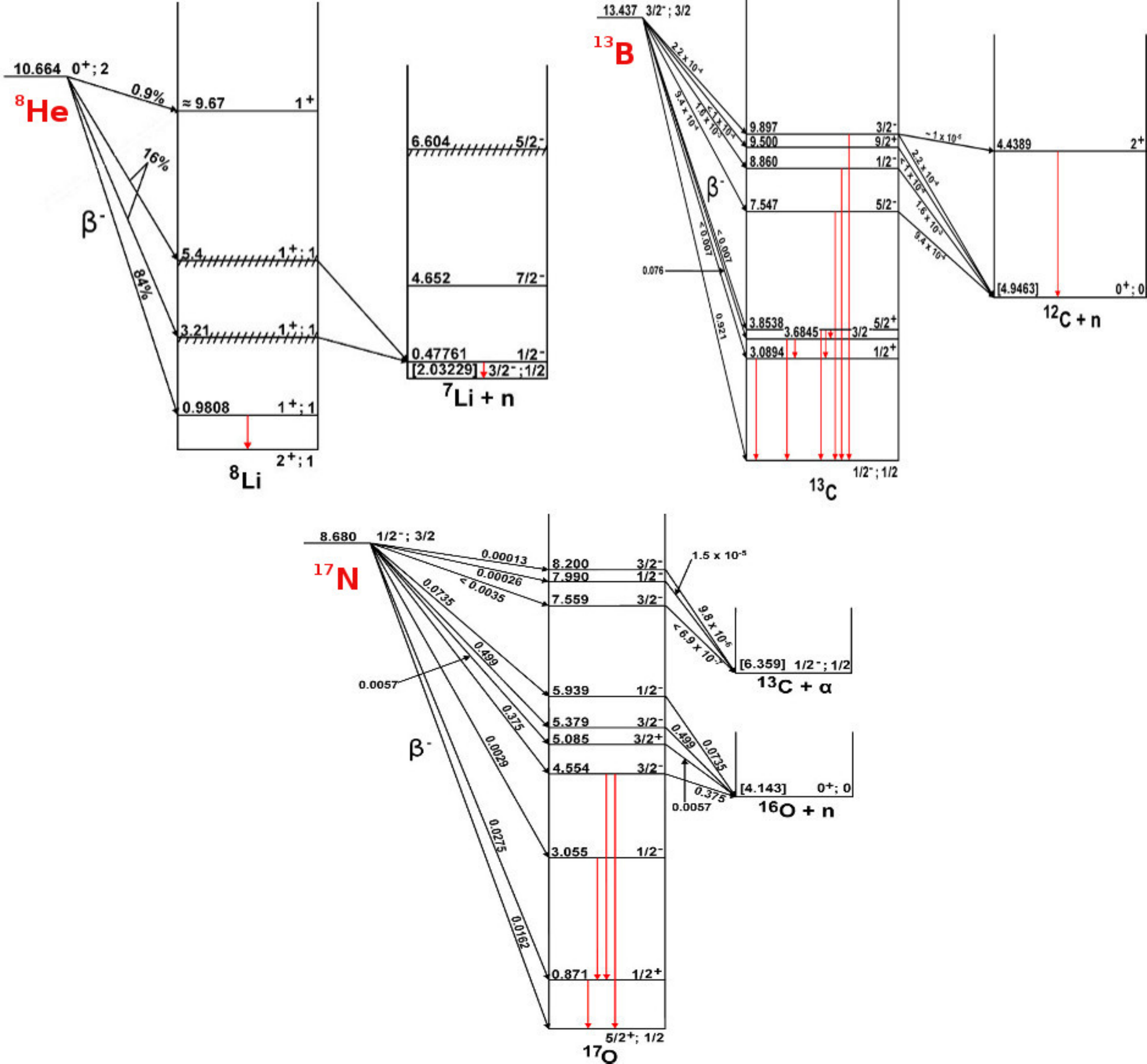}
    \label{fig:spectra}
    \end{figure}

\end{document}